\documentclass[aps,prb,preprint,superscriptaddress,amsmath,amssymb,endfloats*]{revtex4}
\usepackage{graphicx}
\usepackage{dcolumn}
\usepackage{booktabs}
\usepackage{bm}
\usepackage{color}
\usepackage{multirow}
\usepackage{ulem}
\usepackage{setspace}

\newcommand{\Tc}{\ensuremath{T_{\rm c}}}
\newcommand{\gn}{\ensuremath{\gamma_{\rm n}}}

\newcommand{\cel}{\ensuremath{c_{\rm el}}}

\newcommand{\EF}{\ensuremath{E_{\rm F}}}
\newcommand{\kB}{\ensuremath{k_{\rm B}}}
\newcommand{\rxx}{\ensuremath{\rho_{xx}}}
\newcommand{\ryx}{\ensuremath{\rho_{yx}}}

\newcommand{\GIT}{Ge$_{1-x}$In$_{x}$Te}
\newcommand{\SIT}{Sn$_{1-x}$In$_{x}$Te}
\newcommand{\GT}{Ge$_{1-\delta}$Te}

\begin{document}
\title{Evolution of electronic states and emergence of superconductivity in the polar semiconductor GeTe by doping valence-skipping In}
\author{M.~Kriener}
\email[corresponding author: ]{markus.kriener@riken.jp}
\affiliation{RIKEN Center for Emergent Matter Science (CEMS), Wako 351-0198, Japan}
\author{M.~Sakano}
\affiliation{Department of Applied Physics and Quantum-Phase Electronics Center (QPEC), University of Tokyo, Tokyo 113-8656, Japan}
\author{M.~Kamitani}
\affiliation{RIKEN Center for Emergent Matter Science (CEMS), Wako 351-0198, Japan}
\author{M.~S.~Bahramy}
\affiliation{RIKEN Center for Emergent Matter Science (CEMS), Wako 351-0198, Japan}
\affiliation{Department of Applied Physics and Quantum-Phase Electronics Center (QPEC), University of Tokyo, Tokyo 113-8656, Japan}
\author{R.~Yukawa}
\affiliation{Photon Factory, Institute of Materials Structure Science, High Energy Accelerator Research Organization (KEK), Tsukuba, Ibaraki 305-0801, Japan}
\author{K.~Horiba}
\affiliation{Photon Factory, Institute of Materials Structure Science, High Energy Accelerator Research Organization (KEK), Tsukuba, Ibaraki 305-0801, Japan}
\author{H.~Kumigashira}
\affiliation{Photon Factory, Institute of Materials Structure Science, High Energy Accelerator Research Organization (KEK), Tsukuba, Ibaraki 305-0801, Japan}
\affiliation{Institute of Multidisciplinary Research for Advanced Materials (IMRAM), Tohoku University, Sendai 980-8577, Japan}
\author{K.~Ishizaka}
\affiliation{RIKEN Center for Emergent Matter Science (CEMS), Wako 351-0198, Japan}
\affiliation{Department of Applied Physics and Quantum-Phase Electronics Center (QPEC), University of Tokyo, Tokyo 113-8656, Japan}
\author{Y.~Tokura}
\affiliation{RIKEN Center for Emergent Matter Science (CEMS), Wako 351-0198, Japan}
\affiliation{Department of Applied Physics and Quantum-Phase Electronics Center (QPEC), University of Tokyo, Tokyo 113-8656, Japan}
\author{Y.~Taguchi}
\affiliation{RIKEN Center for Emergent Matter Science (CEMS), Wako 351-0198, Japan}

\date{\today}

\begin{abstract}
GeTe is a chemically simple IV\,--\,VI semiconductor which bears a rich plethora of different physical properties induced by doping and external stimuli. These include, among others, ferromagnetism, ferroelectricity, phase-change memory functionality, and comparably large thermoelectric figure of merits. Here we report a superconductor - semiconductor - superconductor transition controlled by finely-tuned In doping. Our results moreover show the existence of a critical doping concentration around $x = 0.12$ in \GIT, where various properties take either an extremum or change their characters: The structure changes from polarly-rhombohedral to cubic, the resistivity sharply increases by orders of magnitude, the type of charge carriers changes from holes to electrons, and the density of states diminishes at the dawn of an emerging superconducting phase. By core-level photoemission spectroscopy we find indications of a change in the In-valence state from In$^{3+}$ to In$^{1+}$ with increasing $x$, suggesting that this system is a new promising playground to probe valence fluctuations and their possible impact on superconductivity. 
\end{abstract}

%\dates{This manuscript was compiled on \today}
%\doi{\url{www.pnas.org/cgi/doi/10.1073/pnas.XXXXXXXXXX}}

%\begin{document}

\maketitle
%\thispagestyle{firststyle}
%\ifthenelse{\boolean{shortarticle}}{\ifthenelse{\boolean{singlecolumn}}{\abscontentformatted}{\abscontent}}{}

%\dropcap{S}
Superconductivity emerges from a wide range of parent materials, including insulators and semiconductors. When charge carriers are doped by partial substitution of one element for another to form out a sufficiently large density of states (DOS) at the Fermi level, superconductivity is established, provided that an effective attractive interaction works among electrons via lattice vibrations. Therefore, choosing appropriate dopant atoms offers to influence the superconductivity through the formation of DOS at the Fermi level, the provision of the attractive interaction among electrons, and the frequency of lattice vibrations. Historically, it was in the early 1960s that Cohen theoretically predicted superconductivity in many-valley semiconductors and semimetals \cite{cohen64a} due to their peculiar band structure, such as GeTe, SnTe, and SrTiO$_3$ \cite{hein64a,schooley65a,hein69a}, which was experimentally confirmed soon after. In particular, SnTe, which has recently regained much attention as a topological crystalline insulator \cite{hsieh12b,ytanaka12b}, exhibits superconductivity below critical temperatures \Tc\ of less than 300~mK. Interestingly, the superconducting transition temperature is strongly enhanced by In doping in its cubic structure \cite{haldolaarachchige16a,kobayashi18a,kriener18a}. To explain this enhancement, the valence-skipping nature \cite{varma88a,dzero05a,hase16a} of the dopant atom In has been often discussed \cite{haldolaarachchige16a,kobayashi18a,kriener18a,hase17a} likewise Bi, Sn, and Tl. In should formally take its divalent state but is expected to form out instead In$^{1+}$ and In$^{3+}$ or a mixture of both. On the basis of the so-called ``negative-$U$ mechanism'' \cite{varma88a}, the valence-skipping nature is predicted to possibly enhance the superconducting interaction as it is discussed for Tl-doped PbTe \cite{matsushita05a}, Ag-doped SnSe \cite{ren13a,wakita17a}, and K-doped BaBiO$_3$ \cite{cava88a}. 

These interesting implications for superconductivity turned our attention to closely related GeTe, which exhibits a rich variety of different physical properties \cite{boschker17a}, such as structural phase change memory functionality \cite{chen86a,lencer08a,xqliu11a} and its magnetic analogue \cite{kriener16a,kriener17a}, ferromagnetism, multiferrocity \cite{cochrane74a,fukuma03c,ftong11a,przybylinska14a,kriegner16a}, and good thermoelectric properties \cite{snyder08a,levin13a,davidow13a} owing to its multi-valley band structure \cite{herman68a,ciucivara06a}. Recently, it has become well known for a large Rashba spin splitting of its bulk bands due to strong spin-orbit coupling and a polar distortion \cite{disante13a,picozzi14a,rinaldi14a,krempasky16a}, as depicted in Fig.~\ref{fig1}a, taking place at about 700~K from cubic ($Fm\bar{3}m$; $\beta$-GeTe) to rhombohedral ($R3m$; $\alpha$-GeTe) accompanied with an elongation of the unit cell along the cubic [111] direction \cite{goldak66a,pawley66a}. The band structure is shown in Figure~\ref{fig1}b for cubic GeTe for the purpose of simplicity. We note that in the case of rhombohedral structure, it has qualitatively the same features, apart from the Rashba spin splitting. The valence band is mainly of Te $5p$ character while the conduction band primarily consists of Ge $4p$. Figure~\ref{fig1}c gives a schematic view of the DOS (left) and the approximate position of the atomic orbitals of the dopant In (right). In both panels the small-gap feature of semiconducting GeTe is apparent (the band gap is of the order of 200 meV at the $L$ point of the Brillouin zone). In reality, however, GeTe features a metallic-like resistivity and superconducts at low temperatures $\Tc < \sim 300$~mK owing to unintentionally doped holes due to Ge deficiency (\GT). Thus far, there have been only a few reports available about the evolution of thermoelectric properties and the structure in \GIT\ \cite{woolley65a,abrikosov76a,lwu17a}. 

In this paper we report the successful synthesis of the whole solid solution \GIT\ by employing a high-pressure synthesis method and the discovery of a doping-induced superconductor -- semiconductor -- superconductor transition by means of transport and specific-heat measurements. At low doping, the resistivity is enhanced by orders of magnitude while the rhombohedral distortion is suppressed. Around $x=0.12$ the system becomes cubic and a new bulk superconducting phase is established at slightly higher doping concentrations. The unit-cell volume shrinks below $x = 0.12$ and starts to expand above with increasing $x$. Coinciding with these transitions, the charge carriers change from hole to electron type. These observations imply that a change of the In-valence states from In$^{3+}$ (electron doping) to In$^{1+}$ (hole doping) may play a role. Core-level photoemission-spectroscopy data support such a scenario, where at higher doping additional features indeed appear, being indicative of the evolution of a different In-valence state. A model based on this valence-state change is proposed and can explain satisfactorily all observed features.

\section*{Results}
%Figure 1
\begin{figure}[t]
\centering
\includegraphics[width=.9\linewidth]{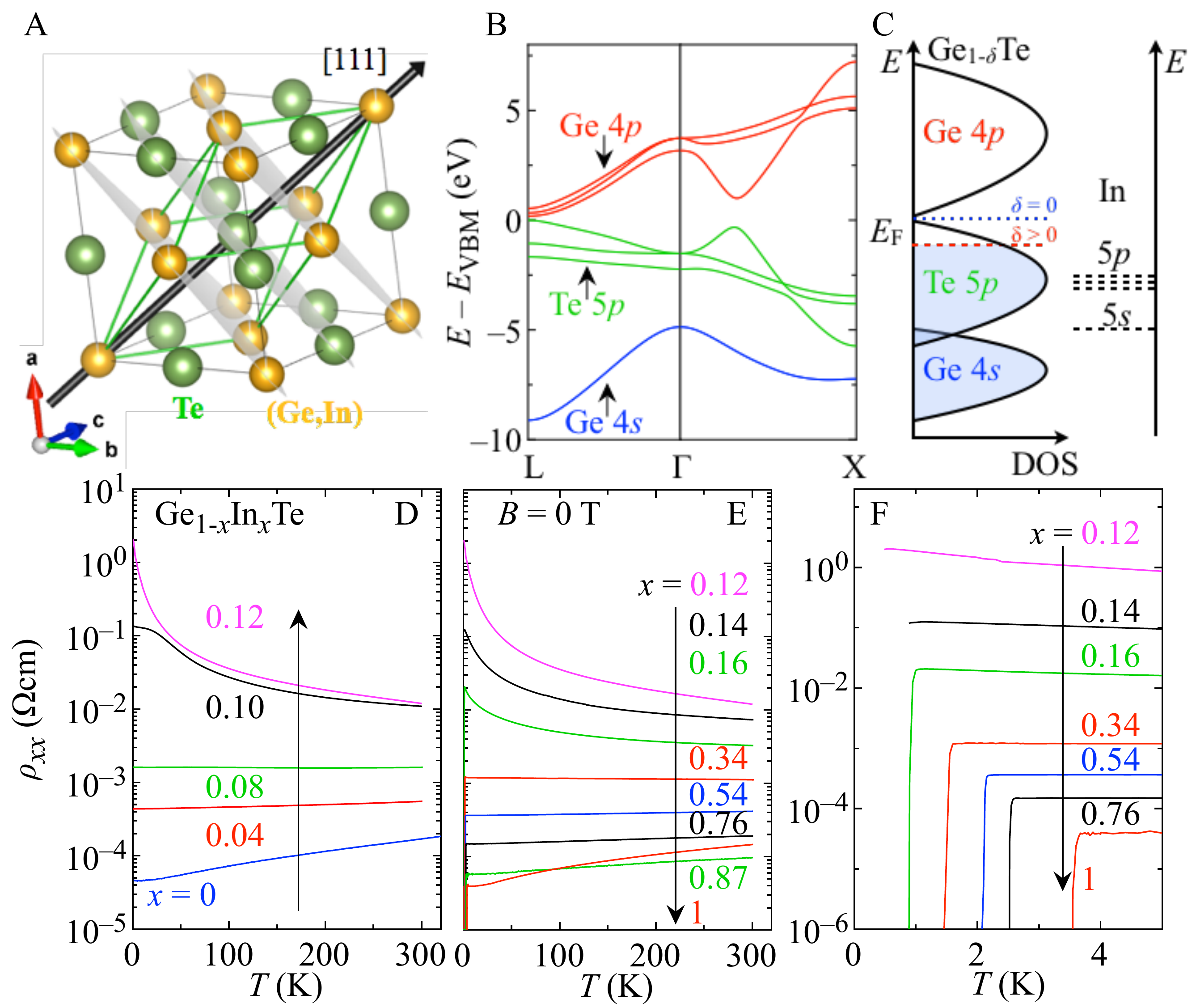}
\caption{(a) Schematic of the unit cell of GeTe. The grey cube denotes the high-temperature cubic unit cell, green the low-temperature rhombohedrally-distorted modification. The black arrow indicates the cubic [111] direction along which the polar distortion takes place, cf.\ Ref.~\cite{kriener16a} for details. (b) Band structure of cubic GeTe. The direct band gap of $\sim 0.2$~eV is located at the $L$ point of the Brillouin zone. The valence-band maximum (VBM) is set to be zero energy. (c) Sketch of the density of states for \GT\ and the atomic energy levels of the In dopant. The blue dotted line indicates the Fermi energy for ideal GeTe without Ge vacancies ($\delta = 0$). The more realistic case of \GT\ is indicated with a red dashed line. (d\,--\,f) Temperature dependence of the longitudinal resistivity of \GIT\ for $0\leq x \leq 1$. (d) Light doping leads to a strong enhancement of the resistivity up to $x = 0.12$. (e) Upon further doping the resistivity is reduced again towards metallic InTe. (f) gives an expanded view for $T\leq 5$~K. Above $x = 0.16$ superconductivity develops as indicated by sharp drops of the resistivity to zero.}
\label{fig1}
\end{figure}
The longitudinal resistivities \rxx\ of selected samples are summarized in Figs.~\ref{fig1}d ($0\leq x \leq 0.12$) and \ref{fig1}e ($0.12\leq x \leq 1$). As for the \GT\ sample used here, we estimate $\delta \approx 1.8$\% from the charge-carrier concentration at room temperature, giving rise to metallic conduction $d\rxx(T)/dT>0$ ($T$: temperature). When doping In, the absolute values of the resistivity increase drastically and the shape of $\rxx(T)$ changes. While $x=0.04$ still exhibits a metallic-like $T$ dependence, this is not the case any more upon higher doping.
Samples with $x\geq 0.1$ exhibit a semiconductor-like $T$ dependence of the resistivity. The largest absolute value of \rxx\ in this study is found for $x = 0.12$, the data of which are shown in both panels (d) and (e) for clarity. As compared to $x=0$, the resistivity at 2~K is enhanced by five orders of magnitude. Nevertheless the absolute value of \rxx\ is still of the order of a few $\Omega$cm and hence cannot be associated with a finite band gap. Upon further increasing $x$, the resistivity becomes again smaller, and for $x > 0.44$ all studied samples exhibit a metallic-like $T$ dependence of \rxx. Figure~\ref{fig1}f provides an expanded view of the low-$T$ region below 5~K, revealing superconducting transitions as indicated by sharp drops of the data for $x \geq 0.16$. Moreover \Tc\ monotonously increases with $x$. 

%Figure 2
\begin{figure}[t]
\centering
\includegraphics[width=.9\linewidth]{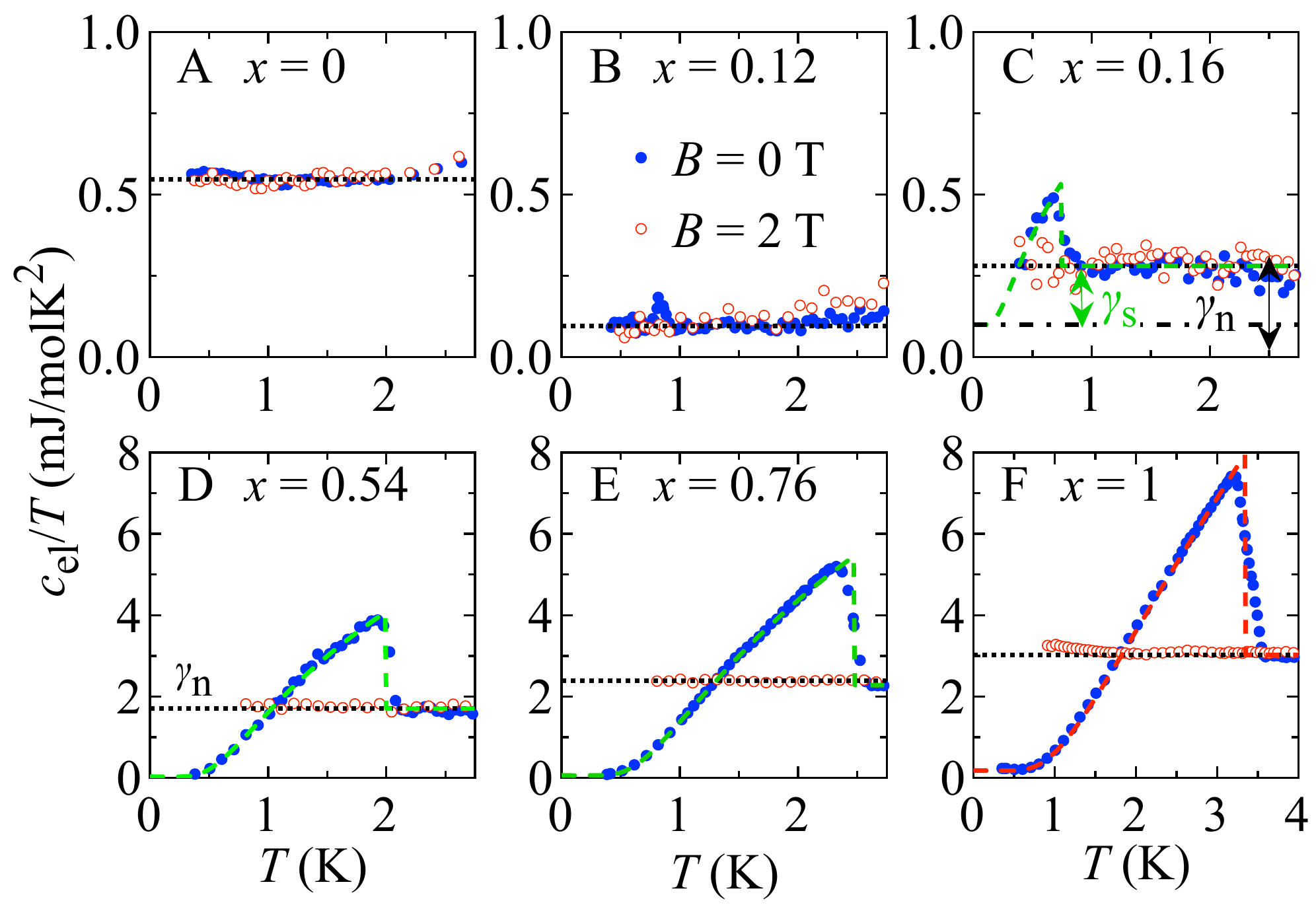}
\caption{Specific heat of \GIT\ for selected $x$ measured in $B=0$~T (blue filled symbols) and $B=2$~T (red open symbols), which is sufficient to suppress the superconductivity. Dotted black lines denote the normal-state electronic specific-heat coefficient \gn, dashed lines represent the electronic specific heat in weak- (green; panels c -- e) or strong-coupling (red; panel f) BCS theory, see text for details. In agreement with resistivity, specific-heat anomalies are observed for $x\geq 0.16$, indicating the formation of a bulk superconducting phase, which does not yet develop over the whole sample in the case of $x = 0.16$ (panel c), as indicated by a residual nonsuperconducting phase with a volume fraction of approximately 35\%, see text. Note the different axes scales for both ordinate and abscissa for each panel.}
\label{fig2}
\end{figure}
Electronic specific-heat data \cel\ of selected samples \GIT\ are displayed as $\cel/T$ vs $T$ plots in Fig.~\ref{fig2}. For the details of the analyses, cf.\ Ref.~\cite{kriener18a}. In agreement with the resistivity results, there is no anomaly visible in data for $x=0$ in the $T$ range $\geq 350$~mK (Fig.~\ref{fig2}a). Doping In leads to a suppression of the normal-state electronic specific-heat coefficient \gn, and hence the DOS at the Fermi level. The lowest \gn\ value is found for a sample with $x=0.12$ (Fig.~\ref{fig2}b) which is most insulating. As already seen in resistivity data, further doping establishes superconductivity in \GIT. For $x = 0.16$ (Fig.~\ref{fig2}c), there is a jump-like anomaly in specific-heat data on top of a residual DOS corresponding to a nonsuperconducting phase fraction. According to our analysis, approximately $ \sim 65$\% of the sample volume superconducts. Upon further doping all samples are found to be bulk superconductors with vanishing or rather small residual DOSs. Moreover, the transitions are sharp, indicating a good sample quality. Up to $x=0.87$, $\cel/T$ data can be well reproduced by weak-coupling BCS theory as shown in Fig.~\ref{fig2}c -- e with $\Delta/\kB\Tc = 1.764$ and $\Delta$ representing the superconducting gap size. However, for $x=1$ it is necessary to increase the BCS coupling strength to 1.95 to yield a satisfactorily description, as shown in Fig.~\ref{fig2}f. This apparent difference is discussed in Section~S8 of the Supporting Information (SI, \cite{Suppl}). As for the samples with $0.12< x < 0.16$, we note that there are drops to zero in resistivity data, but there is no accompanying specific-heat anomaly, indicating filamentary superconductivity.

%Figure 3
\begin{figure}[t]
\centering
\includegraphics[width=.9\linewidth]{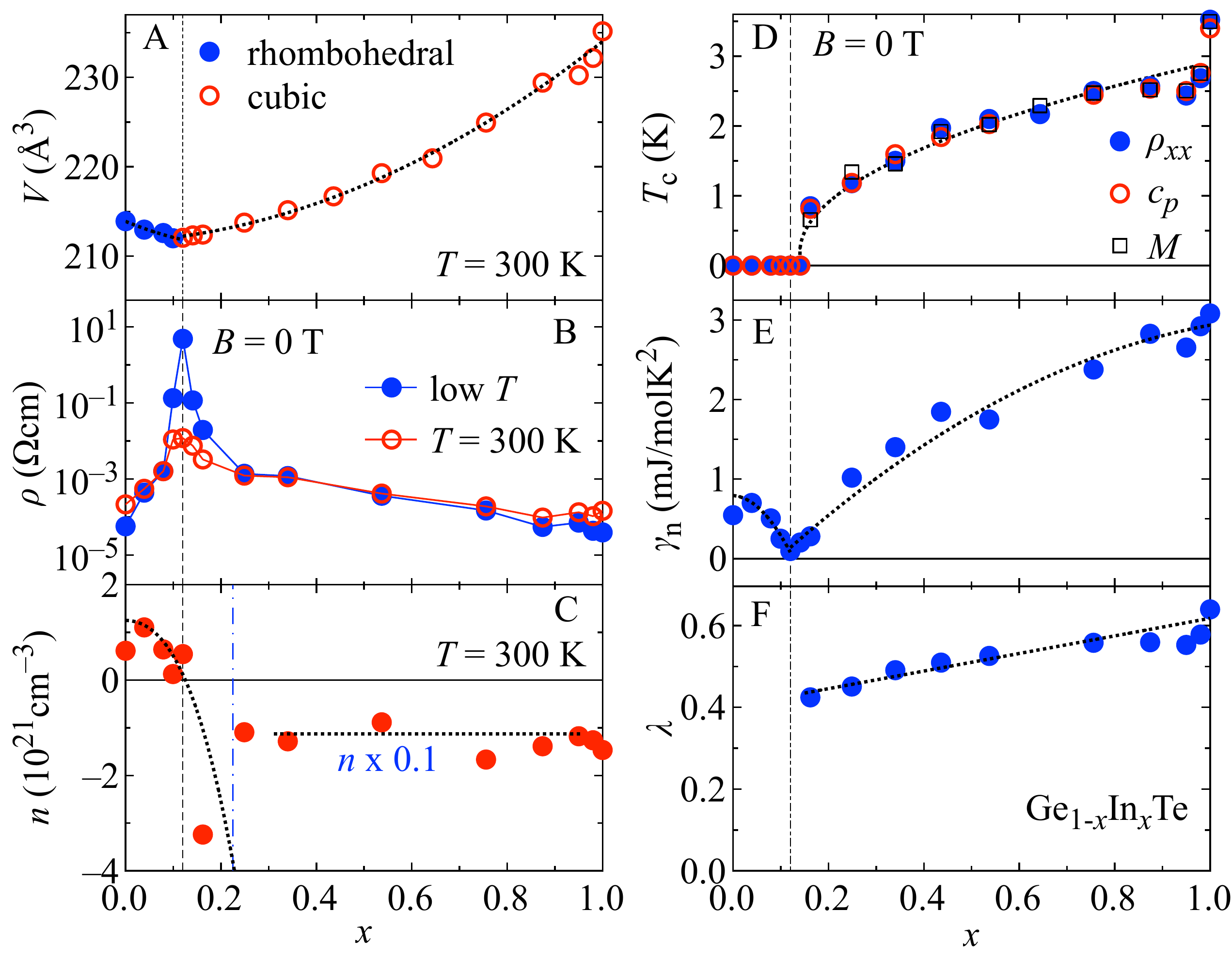}
\caption{Variation of several physical quantities in \GIT. (a) Room-temperature pseudocubic unit-cell volume for rhombohedral structure ($x< 0.12$, blue filled symbols) and cubic one ($x> 0.12$, red open symbols). (b) Zero-field resistivity at room temperature (red open symbols) and at low temperature (at 2~K for $x\leq 0.25$ and above \Tc\ for larger $x$) (blue filled symbols). (c) Room-temperature charge-carrier concentration $n$. Note that $n$ for $x\geq 0.25$ (right of the dashed-dotted vertical line) are multiplied by 0.1 for clarity and found to be roughly constant around $10^{22}$~cm$^{-3}$. (d) Superconducting \Tc\ as estimated from resistivity (filled blue circles), specific heat (open red circles), and magnetization (open black squares). (e) Normal-state electronic specific-heat coefficients. (f) Electron-phonon coupling strength deduced from \Tc\ values and specific-heat data. Dotted lines are guide to the eyes, solid horizontal lines in panels c -- e indicate baselines, and the dashed vertical lines in all panels denote the critical In concentration $x=0.12$.}
\label{fig3}
\end{figure}

Several physical quantities of \GIT\ are summarized in Fig.~\ref{fig3}. The evolution of the unit-cell volume with $x$ is shown in Fig.~\ref{fig3}a. As summarized in Section~S1 of the SI \cite{Suppl}, there is a coexistence region $0.08\leq x < 0.14$ with rhombohedral and cubic phase fractions, and the structure is better described in the rhombohedral $\alpha$-GeTe setting for $x < 0.12$ (blue symbols in Fig.~\ref{fig3}a) and in cubic $\beta$-GeTe above (red open symbols). The most interesting feature here is that the unit-cell volume $V$ shrinks as long as the system is rhombohedrally distorted. By contrast, $V$ strongly increases in the cubic phase. Notably, the overall evolution does not obey Vegard's law, and already above $x \sim 0.25$, the slope of $V(x)$ starts to increase. The $x$ dependence of the corresponding lattice constants are shown in Fig.~S3a of the SI \cite{Suppl}.

Absolute values of the resistivity at room temperature and at low $T$ (at 2~K for $x\leq 0.25$ and above \Tc\ for larger $x$) are plotted against $x$ in Fig.~\ref{fig3}b. The sharp and strong enhancement of \rxx\ around $x\sim 0.12$ is most pronounced at low $T$ and still clearly recognized at 300~K, highlighting this critical In-doping concentration in \GIT.

In Fig.~\ref{fig3}c the charge-carrier concentrations $n$ are plotted against $x$ as estimated from magnetic-field $B$ dependent Hall-effect $\ryx(B)$ measurements at room temperature, although $n$ deduced from $\ryx(B)$ may show some deviation from the real carrier concentration for metallic samples. The hole-type charge-carrier concentration is quickly suppressed when introducing In. The resulting charge-neutrality point is located around $x = 0.12$, i.e., the most insulating doping range. In spite of the semiconductor-like slope of $\rxx(T)$ for $0.25 \leq x < 0.44$, the electron concentrations in these samples are already of the order of $10^{22}$~cm$^{-3}$ and hence the conduction regime is barely metallic. For $x>0.44$, $n$ stays almost constant around $10^{22}$~cm$^{-3}$. The cation deficiency $\delta$, which may affect the carrier density, is examined by a scanning electron microscope equipped with an energy-dispersive x-ray (SEM-EDX) analyzer on selected samples as described in Section~S5 of the SI \cite{Suppl}.

Superconducting \Tc\ values as estimated from resistivity, specific-heat, and magnetization data increase monotonously with $x$ and agree well with each other, see Fig.~\ref{fig3}d. Interestingly, near InTe, \Tc\ increases very rapidly. 

Figure~\ref{fig3}e shows the evolution of \gn\ with $x$. The \gn\ value of the measured GeTe sample has a smaller value than the sample for $x=0.04$, probably due to the particular value of the Ge deficiency of the examined specimen. Upon increasing the In concentration, \gn\ is reduced and almost zero but finite around $x=0.12$ as already suggested by the evolution of the charge-carrier concentration. For larger $x$, \gn\ increases monotonously.  

The final panel Fig.~\ref{fig3}f summarizes the electron-phonon coupling strength $\lambda$ as estimated from our quantitative specific-heat analyses. It increases with almost constant slope over the superconducting doping range $0.16\leq x \leq 1$. Interestingly the strong enhancement of \Tc\ for $x=1$ only is neither clearly reflected in \gn\ nor in $\lambda$, cf.\ also Section~S8 in the SI \cite{Suppl}.

%Figure 4
\begin{figure}[t]
\centering
\includegraphics[width=.9\linewidth]{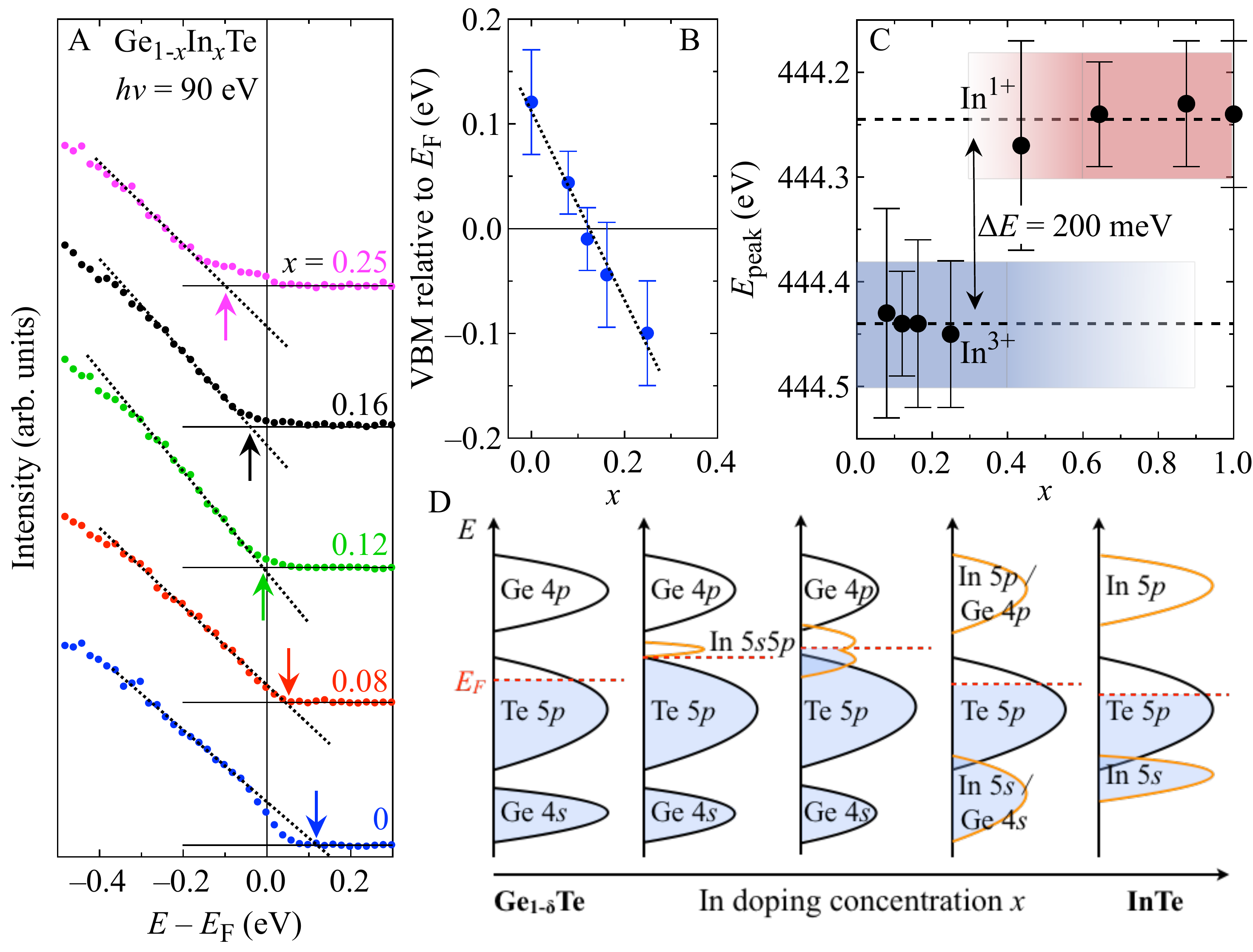}
\caption{(a) Valence-band photoemission spectra of \GIT\ for $x \leq 0.25$ recorded by the photon energy of $h\nu = 90$~eV. Data for different $x$ are shifted with respect to each other for clarity. The vertical solid line represents \EF. At each data set, horizontal solid lines indicate the baseline and dotted lines are linear fits to the data below \EF. The arrows denote the energies where the solid and dotted lines intersect for each $x$, respectively, as a measure of the valence-band-maximum (VBM) energy. (b) Replotted VBM energies as a function of $x$. The sign change indicates where \EF\ shifts above the VBM, coinciding with the critical In concentration $x = 0.12$. The dotted line is a guide to the eyes. (c) Peak energy of In-$3d_{5/2}$-core-level photoemission spectra of selected \GIT\ samples is plotted against $x$, demonstrating the two different In-valence states. Gradations indicate the coexistence region of both valence states, and the change from mainly In$^{3+}$ (blue; low doping) to mainly In$^{1+}$ (red; high doping). Dashed horizontal lines indicate the average peak energy of each feature which differ by approximately 200~meV. (d) Schematic illustration of the evolution of the band structure in \GIT\ with $x$, see text for details.}
\label{fig4}
\end{figure}
To obtain information on how the electronic structure changes upon In doping, we performed photoemission spectroscopy. Figure~\ref{fig4}a shows the valence-band spectra for $x\leq 0.25$ around the Fermi energy \EF, which is indicated by a vertical solid line. The observed behavior is typical for a $p$-type semiconducting system. Arrows indicate the valence-band maximum (VBM) energy relative to \EF, defined as intersection point of both solid and dotted lines for each $x$. The energy values of the VBM are replotted in Fig.~\ref{fig4}b. Apparently the VBM shifts linearly with $x$ from above to below \EF\ and coincides with \EF\ at the critical In concentration $x=0.12$, indicating the depletion of the charge carriers at, and their sign change across this doping level. At higher doping $x\geq 0.44$, the spectra change qualitatively from semiconducting to metallic as can be seen in Figs.~S6a and b of the SI \cite{Suppl}. A step or edge at \EF\ reflects the metallic character of these samples. 

The bulk-sensitive x-ray PES measurements ($h\nu = 1486.6$~eV) for the In-$3d_{5/2}$ core-level allowed us to obtain information about the valence state of In, cf.\ Fig.~S6 of the SI \cite{Suppl}. In the intermediate $x$ region of $0.25 < x < 0.64$, the core-level structure broadens, and the peak position changes suddenly around $x=0.34$. The peak energies are replotted in Fig.~\ref{fig4}c. These two values are associated with the two valence states of In. Dashed horizontal lines are guides for the energies of both features. They are separated by approximately 0.2~eV, similarly as the Sn$^{2+}$ and Sn$^{4+}$ peaks in Sn core-level spectra of AgSnSe$_2$ \cite{wakita17a} or Sn oxides \cite{themlin92a}. In the case of AgSnSe$_2$, Sn$^{2+}$ and Sn$^{4+}$ peaks appear at the binding energies of $\sim 485.6$ and $\sim 486.3$~eV, respectively, thus  indicating a separation of $\sim 0.7$~eV. In analogy with this behavior, we attribute the energetically shallower feature ($\sim 444.25$~eV) to the In$^{1+}$ and the deeper one ($\sim 444.44$~eV) to the In$^{3+}$ state.

\section*{Discussion}
Apparently $x_{\rm c} = 0.12$ is a critical In concentration around which several properties of \GIT\ change dramatically. The structure changes from rhombohedral to cubic, the unit cell shrinks below and expands above, the resistivity is strongly enhanced by five orders of magnitude within a very small doping range around $x_{\rm c}$, the charge-carrier type changes from hole to electron, superconductivity emerges, and the DOS is depleted. Given that \GT\ is a very-low-\Tc\ superconductor, the system apparently runs through a superconductor -- semiconductor -- superconductor transition. Also, as described in the introduction, the In dopant is a so-called valence-skipping element with favorable In$^{1+}$ ($4d^{10}5s^2$) and In$^{3+}$ ($4d^{10}5s^0$) valence states \cite{varma88a,erickson09a,hase16a}. Therefore it is reasonable to think about their role in this system. 

The sketch in Fig.~\ref{fig4}d illustrates the plausible evolution of the In states as a function of the In concentration on the basis of the results summarized in Figs.~\ref{fig1}\,--\,\ref{fig4}. The left-most schematic DOS shows the situation in \GT\ and is similar to that in Fig.~\ref{fig1}c. The Fermi level (red dashed line) lies in the Te-$5p$ band, giving rise to metallic-like conduction with a hole-type carrier concentration of the order of $10^{21}$~cm$^{-3}$. The second schematic picture shows the situation for light In doping, which very effectively reduces the hole-type carriers, shifting the Fermi level upwards. From literature \cite{haldolaarachchige16a} it is known that light In doping leads to the formation of impurity states located at the top of the VBM. This is also confirmed by our band calculation for $x=0.12$ (see Section~S9 in the SI \cite{Suppl}). In the sketch, this feature is labelled ``In $5s5p$'' to emphasize its origin from the respective atomic In orbitals. These newly-formed states are mostly empty, and hence the valence state of In is 3+. Upon further doping, the impurity band becomes wider and the initial ``In $5s5p$'' states start to separate as shown in the central schematic drawing of Fig.~\ref{fig4}d, and the conduction mechanism will gain again metallic character above $x = 0.12$. The next schematic shows how the bands of In $5p$\,--\,Ge $4p$ and In $5s$\,--\,Ge $4s$ characters form mixed orbital states at higher doping, called ``amalgamated bands'' in Ref.~\cite{onodera68a}. When the doping level is sufficiently high, the In-$5s$ orbitals will have developed into a proper fully occupied band which has shifted below the Te-$5p$ band and hence well below \EF. Only the In-$5p$ band remains empty, thus In now takes its 1+ state.

The analysis of the In-$3d_{5/2}$ core-level photoemission spectra allows us to further confirm this: For low doping up to approximately $x=0.25$, we can only identify the feature at higher binding energy (In$^{3+}$). From $x=0.25$ a second feature develops, indicating that the valence state of the dopants start to become 1+. At the same time the peak indicative of the In$^{3+}$ state fades away and is hardly seen for $x > 0.44$, cf.\ Fig.~S6c of the SI \cite{Suppl}. Therefore the crossover from 3+ to 1+ mainly takes place in the intermediate In-doping range between $\sim 0.25$ and $\sim 0.64$, cf.\ Fig.~\ref{fig4}c. The change in the In valence states is also reflected in the nonmonotonous behavior of the unit-cell volume $V$ with $x$, cf.\ Fig.~\ref{fig3}a: $V$ decreases as $x$ increases to $x_{\rm c}=0.12$ because of the smaller ionic radius of In$^{3+}$, but tends to nonlinearly increase above $x_{\rm c}$ due to the increasing fraction of In$^{1+}$ ions with larger radius. At higher doping, the system behaves like a simple metal as also indicated by resistivity data shown in Fig.~\ref{fig1}(e). This situation is sketched in the final drawing in Fig.~\ref{fig4}d for pure InTe, where a metallic ground state with a large Fermi surface and empty In-$5p$ bands is realized. To assume the 1+ state for In is reasonable even for metallic InTe (without ionic bonds), because of the energetic proximity of the Te $5p$ and In $5p$ states which allows for an easy charge transfer between them.

Finally we discuss some striking differences between In-doped SnTe and In-doped GeTe, given that both are isostructural and share qualitatively many similar features. Charge carriers originate from unintentionally self-doped holes leading to low-temperature superconductivity, which is depleted by doping In. However, when doping In as little as $x \sim 0.017$, \SIT\ becomes again superconducting with \Tc\ values of up to $\sim 2$~K \cite{erickson09a,novak13a}. Since the suppression of the polar distortion requires In doping of $x \sim 0.04$, lower-doped \SIT\ is a polar superconductor. By contrast, in \GIT\ the doping-induced superconducting phase only emerges above the structural transition $x> 0.12$. While in \GIT\ the charge neutrality point and hence the change of the carrier type coincides with the structural transition and various other features, in \SIT\ the sign change of the carriers happens around $x\sim 0.08 - 0.1$ \cite{haldolaarachchige16a} which is higher than the concentration of the structural transition. Another significant difference is the $x$ dependence of \Tc: \SIT\ features a two-dome phase diagram with a strong and sudden suppression of \Tc\ around $x=0.58$ \cite{kriener18a}. By contrast, \Tc\ monotonously increases with $x$ in \GIT, as shown in Fig.~\ref{fig3}d. This might be a consequence of the difference in the respective ionic radii and atomic energy levels between Ge and Sn, giving rise to distinct degrees of local lattice distortion and hybridization with the In and Te states. However, the band inversion apparent in SnTe, which was reported to survive at least light In doping \cite{ytanaka12b,tsato13a}, may further complicate the situation. To chase down the origins of these pronounced differences between the two systems \GIT\ and \SIT\ that are at first glance very similar, and to answer the question about the exact role of the valence state of In are promising starting points for future studies.
\vspace{1cm}

\section*{Acknowledgments}
This work was partly supported by Grants-In-Aid for Scientific Research (S) from the Japan Society for the Promotion of Science (JSPS, No.\ 24224009), JST (No.\ JP16H00924), and PRESTO (JPMJPR15N5) and Grants-In-Aid for Scientific Research (B) (JSPS, No.\ 17H02770).  MK is supported by a Grants-in-Aid for Scientific Research (C) (JSPS, No.\ 15K05140). We thank R.~Arita and T.~Koretsune for fruitful discussions.
%}

%\showacknow{}

% Bibliography

%\bibliography{/PaperBase/Bibfiles/preload,/PaperBase/Bibfiles/AB-semiconductors,/PaperBase/Bibfiles/Thermoelectrics,/PaperBase/Bibfiles/Superconductivity,/PaperBase/Bibfiles/sonstigePaper,/PaperBase/Bibfiles/Lehrbuch,/PaperBase/Bibfiles/TopolIns,/PaperBase/Bibfiles/SiC-Si-C,additionalbib}

\end{document}

\title{Supporting Information for \\ ``Evolution of electronic states and emergence of superconductivity in the polar semiconductor GeTe by doping valence-skipping In''}

\author{M.~Kriener}
\email[corresponding author: ]{markus.kriener@riken.jp}
\affiliation{RIKEN Center for Emergent Matter Science (CEMS), Wako 351-0198, Japan}
\author{M.~Sakano}
\affiliation{Department of Applied Physics and Quantum-Phase Electronics Center (QPEC), University of Tokyo, Tokyo 113-8656, Japan}
\author{M.~Kamitani}
\affiliation{RIKEN Center for Emergent Matter Science (CEMS), Wako 351-0198, Japan}
\author{M.~S.~Bahramy}
\affiliation{RIKEN Center for Emergent Matter Science (CEMS), Wako 351-0198, Japan}
\affiliation{Department of Applied Physics and Quantum-Phase Electronics Center (QPEC), University of Tokyo, Tokyo 113-8656, Japan}
\author{R.~Yukawa}
\affiliation{Photon Factory, Institute of Materials Structure Science, High Energy Accelerator Research Organization (KEK), Tsukuba, Ibaraki 305-0801, Japan}
\author{K.~Horiba}
\affiliation{Photon Factory, Institute of Materials Structure Science, High Energy Accelerator Research Organization (KEK), Tsukuba, Ibaraki 305-0801, Japan}
\author{H.~Kumigashira}
\affiliation{Photon Factory, Institute of Materials Structure Science, High Energy Accelerator Research Organization (KEK), Tsukuba, Ibaraki 305-0801, Japan}
\affiliation{Institute of Multidisciplinary Research for Advanced Materials (IMRAM), Tohoku University, Sendai 980-8577, Japan}
\author{K.~Ishizaka}
\affiliation{RIKEN Center for Emergent Matter Science (CEMS), Wako 351-0198, Japan}
\affiliation{Department of Applied Physics and Quantum-Phase Electronics Center (QPEC), University of Tokyo, Tokyo 113-8656, Japan}
\author{Y.~Tokura}
\affiliation{RIKEN Center for Emergent Matter Science (CEMS), Wako 351-0198, Japan}
\affiliation{Department of Applied Physics and Quantum-Phase Electronics Center (QPEC), University of Tokyo, Tokyo 113-8656, Japan}
\author{Y.~Taguchi}
\affiliation{RIKEN Center for Emergent Matter Science (CEMS), Wako 351-0198, Japan}

\date{\today}

\maketitle

\subsection{Crystal Structure at room temperature}
\label{xrd}
\begin{figure}[h]
\centering
\includegraphics[width=15cm,clip]{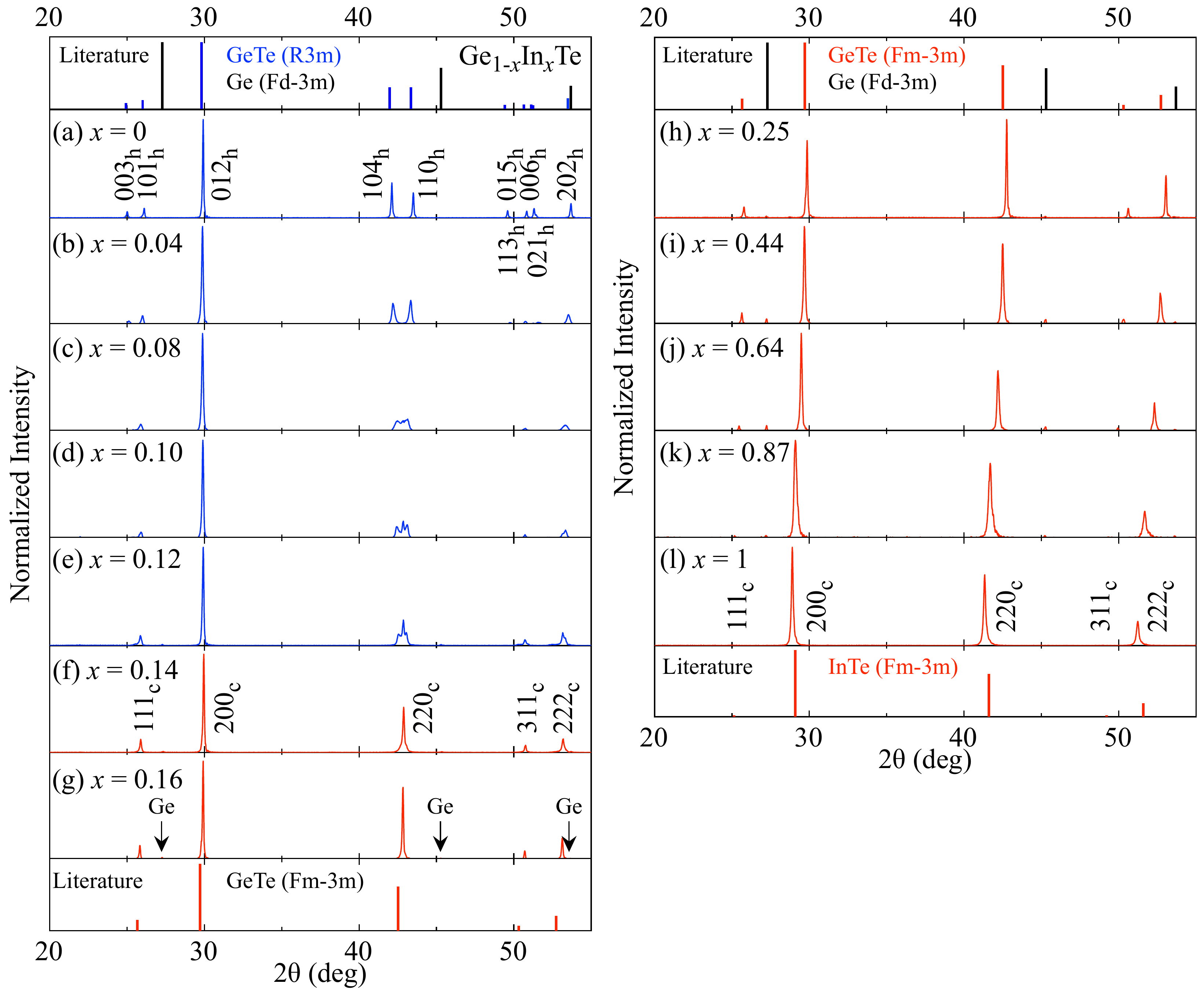}
\caption{XRD data of selected batches of \GIT. The respective peaks can be indexed in hexagonal and cubic settings for $x< 0.12$ and $x>0.12$, respectively. Arrows in panel (g) denote the position of tiny impurity peaks due to the presence of unreacted cubic Ge, see text. The upper- and lowermost panels show the expected peak positions and their intensities for the relevant compounds.}
\label{figS1}
\end{figure}

Figure~\ref{figS1} summarizes x-ray diffraction (XRD) patterns of various batches of \GIT\ ($0\leq x \leq 1$). The intensities in all panels are normalized with respect to each main peak. The peaks of \GIT\ ($x<0.14$) are indexed in the hexagonal setting of the rhombohedral unit cell, and the reflections for $x\geq 0.14$ in cubic symmetry. Arrows in panel (g) denote the position of impurity peaks due to the presence of tiny amounts of unreacted cubic Ge, which seems unavoidable in agreement with literature.\cite{woolley65a} The structural transition takes place in the In doping range $0.04 < x < 0.14$. This can be best monitored when looking at the $104_{\rm h}$ and $110_{\rm h}$ reflections in the rhombohedral phase which merge into the single $220_{\rm c}$ reflection in the cubic phase. At higher doping levels the peaks clearly shift towards their cubic InTe counterparts, indicating substitutional doping as also seen in the evolution of lattice constant and unit-cell volume in Fig.~\ref{figS3}.
%Figure S1

\clearpage

\subsection{Time and temperature dependence of the crystal structure in \GIT}
\label{xrdtime}
\begin{figure}[h]
\centering
\includegraphics[width=15cm,clip]{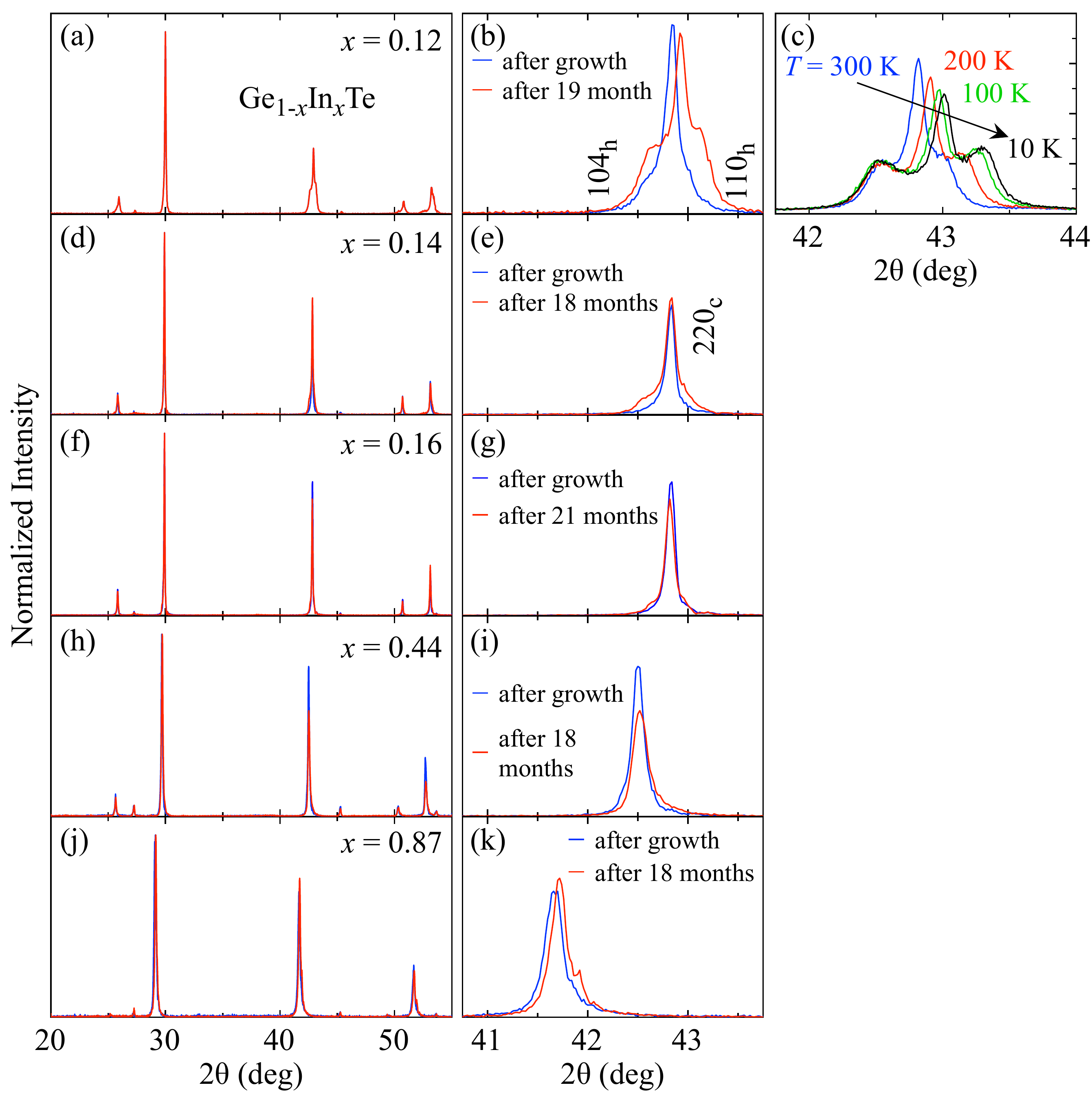}
\caption{Aging effect on XRD data of selected batches \GIT. Blue data in panels (a), (b), and (d) -- (i) were measured right after the crystal growth had finished. Red data were measured after more than 1.5 years had passed on the same powder which was kept at room temperature.  Panel (c) summarizes temperature-dependent XRD data for $x=0.12$ taken upon cooling from room temperature, see text for details.}
\label{figS2}
\end{figure}

In literature it was reported that keeping \GIT\ samples at elevated temperatures leads to a dissociation,\cite{woolley65a} which we confirmed: When \GIT\ samples experience elevated temperatures of a few hundred $^{\circ}$C, the samples decompose irreversibly into a multiphase mixture. Pure cubic InTe was reported to switch back into its ambient-pressure tetragonal modification at a speed of 20\% in four months.\cite{banus63a} Therefore we checked the long-time stability of our samples by remeasuring XRD of selected powder samples, which were initially measured right after the crystal growth had finished. The results are summarized in Fig.~\ref{figS2}. The central panels provide an expanded view of the rhombohedral $104_{\rm h}$ and $110_{\rm h}$ or cubic $220_{\rm c}$ reflections, where the difference between the two structural modifications is best seen. The higher doping concentrations $x\geq 0.14$ are not much affected by aging, i.e., the meta-stable cubic structure is kept even in powder samples which were stored at room temperature. For samples with In concentrations in the structural transition range, the aging effect is stronger although the overall structural situation remains unchanged: For $x =0.12$ the splitting of the cubic $220_{\rm c}$ peak towards rhombohedral structure has increased slightly over time. For this In concentration we also measured XRD patterns below room temperature, which are summarized in panel (c). Upon decreasing temperature, the rhombohedral distortion is enhanced while the intensity of the cubic phase decreases. The observed behavior suggests that the structural transition is of first order in nature.
%Figure S2

\clearpage

\subsection{Employing different high-pressure synthesis recipes}
\label{grrec}
Motivated by the apparent differences in the superconducting \Tc\ of pure InTe reported in literature,\cite{kobayashi18a,kriener18a,boemmel63a,banus63a,geller64a,geller64b,hulm70a} 
we studied this in more detail. We found that the growth temperature in our high-pressure synthesis approach has impact on several physical properties of \GIT\ for $x > 0.87$, most drastically in pure InTe as summarized in Table~\ref{tab1} in Section~\ref{SCInTe}. In agreement with an early publication,\cite{geller64a} our SEM-EDX analyses showed that this is correlated with deviations in the In concentration: \IT. In some of the following subsections, we present results on In-rich batches grown according to two different high-pressure recipes referred to as recipe A (high-pressure synthesis at 5~GPa, $1200^{\circ}$C, 1~h) and B (5~GPa, $600^{\circ}$C, 1~h). 

\subsection{Doping dependence of the unit-cell volume}
\label{ucv}
\begin{figure}[h]
\centering
\includegraphics[width=8cm,clip]{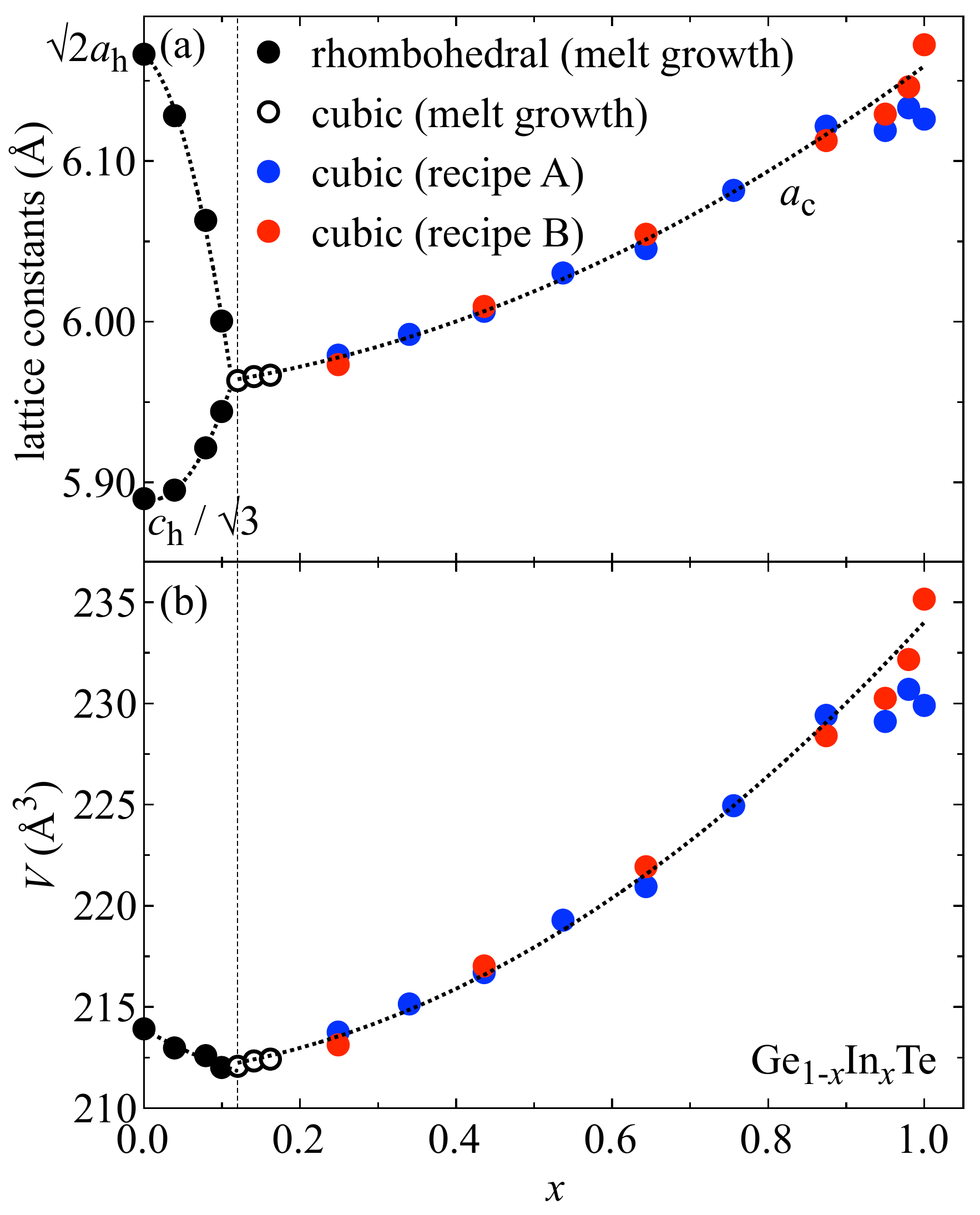}
\caption{Doping dependence of (a) lattice constants and (b) unit-cell volume. The vertical solid line indicates the critical In concentration $x_{\rm c}=0.12$. Dotted lines are guides to the eyes. Black data points refer to batches which were synthesized by conventional melt growth, blue and red symbols to batches synthesized according to recipes A and B, respectively, see text.}
\label{figS3}
\end{figure}

Figure~\ref{figS3} shows (a) lattice constants and (b) unit-cell volume of \GIT\ vs $x$. The unit-cell volume is already shown in Fig.~3(a) of the main text. However, the color code of Fig.~\ref{figS3} is different as compared to Fig.~3(a): Black data points refer to melt-grown batches, blue and red symbols indicate whether the corresponding batches were grown according to high-pressure recipe A or B, respectively, as defined in Section~\ref{grrec}. While for $x\leq0.87$ both recipes yield samples with similar lattice constants, this changes upon further increasing the In concentration. Samples grown according to recipe B, i.e., at lower temperatures, exhibit larger lattice constants than those grown at higher temperatures (recipe A). This enhancement effect is already seen for $x = 0.95$ and becomes stronger when approaching $x = 1$. For the latter the difference in the cubic lattice constant is as large as $\sim 0.7$\%, cf.\ Section~\ref{SCInTe} and Table~\ref{tab1}.

As summarized in Section~\ref{xrd}, there is a coexistence region $0.08\leq x < 0.14$ with rhombohedral and cubic phase fractions, and the structure is better described in the rhombohedral $\alpha$-GeTe setting for $x < 0.12$ and in cubic $\beta$-GeTe above. Therefore Fig.~\ref{figS3}(a) shows two lattice constants for $x < 0.12$ in pseudocubic setting $a_{\rm c} = \sqrt{2}a_{\rm h}$ and $c_{\rm c} = c_{\rm h}/\sqrt{3}$ with $a_{\rm h}$ and $c_{\rm h}$ denoting the corresponding lattice constants in hexagonal notation, and a single cubic $a_{\rm c}$ above $x=0.12$.
 
%Figure S3

\clearpage

\subsection{SEM-EDX analysis}
\label{edx}
\begin{figure}[h]
\centering
\includegraphics[width=8cm,clip]{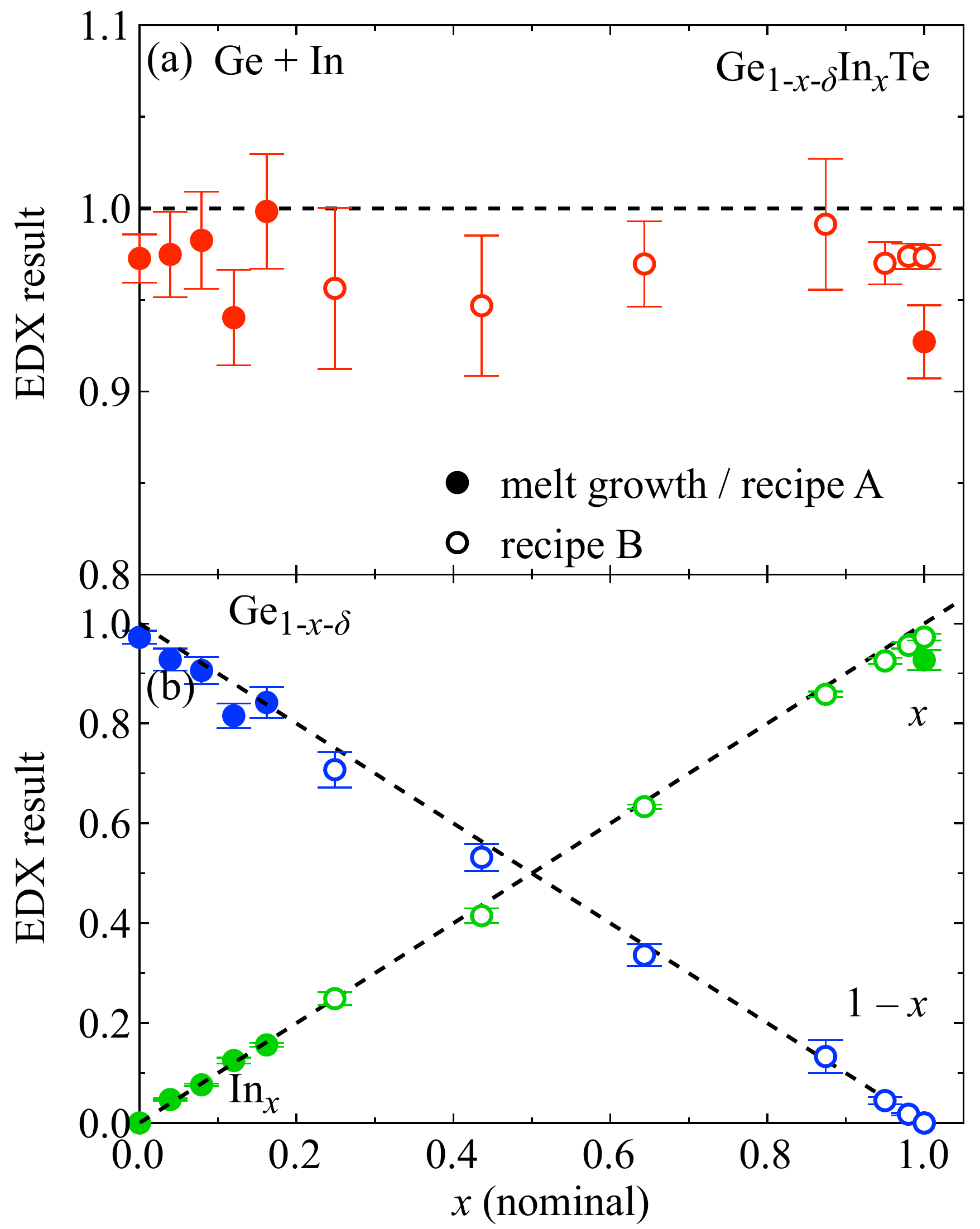}
\caption{Result of SEM-EDX analyses of various batches \GIT. Filled symbols refer to batches synthesized by melt growth or according to high-pressure recipe A, open data points to recipe B. Ge hot spots are excluded from the analyses. In (a) the sum of Ge and In counts is shown which slightly deviates from unity (dashed line) even for $x>0.12$ due to Ge clustering in the samples, see text. (b) In and Ge counts. The dashed lines indicate the nominal In and Ge concentrations.}
\label{figS4}
\end{figure}

Figure~\ref{figS4} summarizes results obtained by employing a scanning-electron microscope equipped with an energy-dispersive x-ray (SEM-EDX) analyzer. For each sample usually two to three different larger surface areas of several hundred~$\mu$m$^2$ were selected and within each area four to six smaller areas analyzed quantitatively by EDX. Obtained Ge and In contents are normalized by the Te content. Then the results were averaged and the standard deviations calculated which define the error bars in Fig.~\ref{figS4}. Filled symbols refer to melt growth / high-pressure recipe A, open symbols to recipe B. 

As already suggested by our XRD results (cf.\ Fig.~\ref{figS1}), SEM-EDX data reveals the existence of Ge clusters for $x\geq 0.04$ and hence there is some Ge inhomogeneity. These clusters have a typical size of $\sim \mu$m$^2$. For the EDX analyses, we excluded those ``Ge hot spots'', leading to slight deviations of the ``Ge + In'' count from unity even above $x = 0.12$, for which this effect seems strongest. The In concentration is very close to the nominal one for $x\leq 0.87$ as shown in Fig.~\ref{figS4}(b) by the dashed line labeled as ``$x$''. Above, the deviation slightly (strongly) increases for batches grown according to recipe B (A). Nevertheless In is distributed homogeneously in the whole doping series as indicated by small error bars and SEM images (not shown). The Ge concentration slightly deviates from the expected ``$1-x$'' line due to the clustering.  Only at such ``Ge hot spots'', In and Te are lacking. 
%Figure S4

\clearpage

\subsection{Magnetization}
\label{Mag}
\begin{figure}[h]
\centering
\includegraphics[width=12cm,clip]{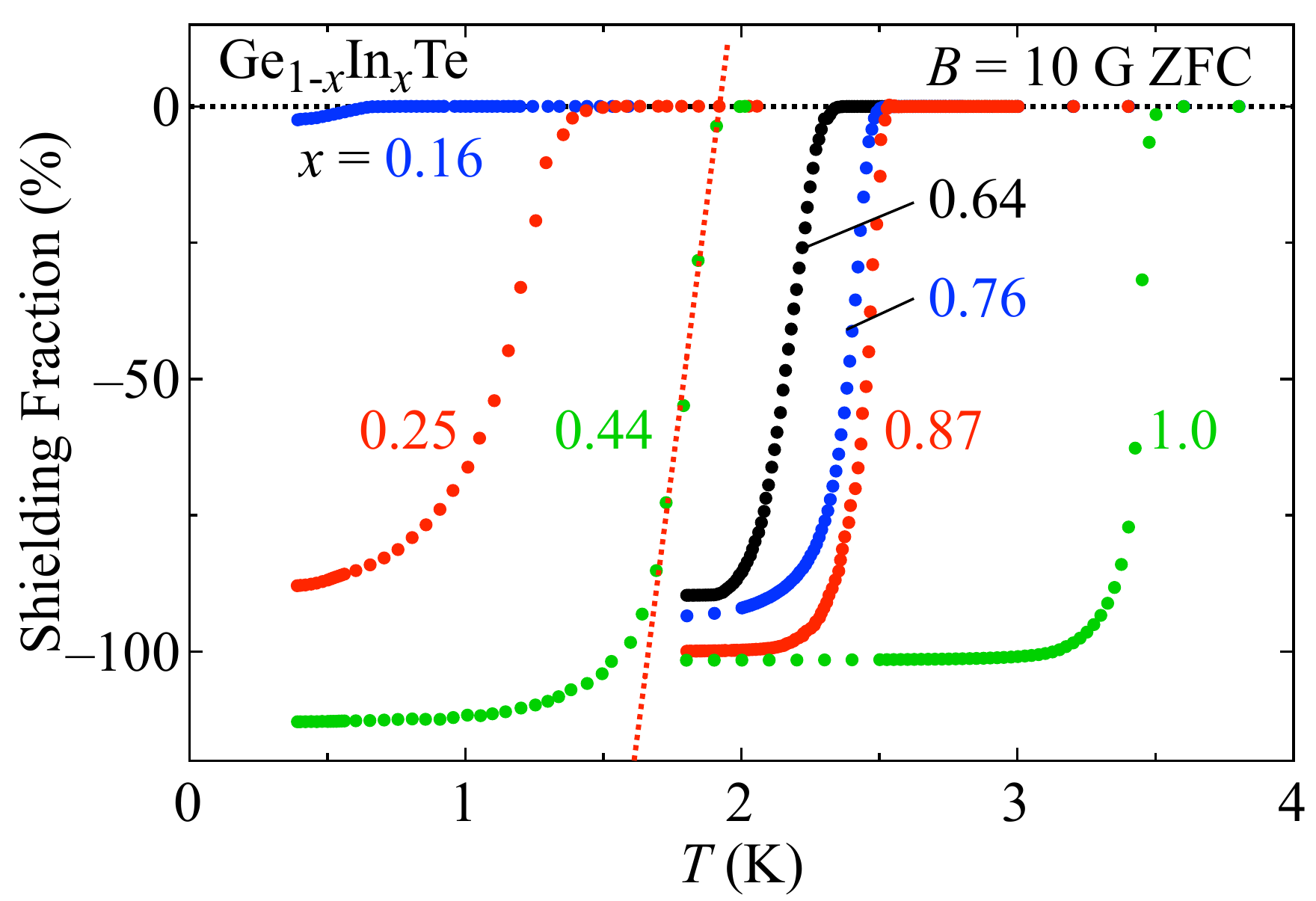}
\caption{Superconducting shielding fraction of \GIT. The demagnetization effect is appropriately corrected for (see text). The linear fit (red-dotted line) to the data for $x=0.44$ shows how \Tc\ is defined.}
\label{figS5}
\end{figure}

Figure~\ref{figS5} shows the superconducting shielding fraction for various samples of \GIT\ as estimated from temperature-dependent magnetization data. The data has been corrected for the demagnetization effect by employing Brandt's formula for samples in slab-like geometry.\cite{brandt99a} According to specific-heat data, all samples except $x = 0.16$ exhibit more than $\sim 85$\% volume fraction. In magnetization most samples agree with this result within some error bars: Deviations from 100\% in magnetization data are likely due to problems in determining the exact field strength because the remanent field of the superconducting magnet used is not precisly known, deviations from exact slab-like geometry, and the estimation of the sample volume and linear dimensions needed for the demagnetization-effect correction. The linear fit to the data for $x=0.44$ shows exemplarily how \Tc\ is defined in the case of magnetization data: The temperature at which the fit for each sample intersects with the dotted baseline is plotted in the phase diagram in Fig.~3d of the main text.
%Figure S5

\clearpage

\subsection{Photoemission spectroscopy}
\label{PES}
\begin{figure}[h]
\centering
\includegraphics[width=14cm,clip]{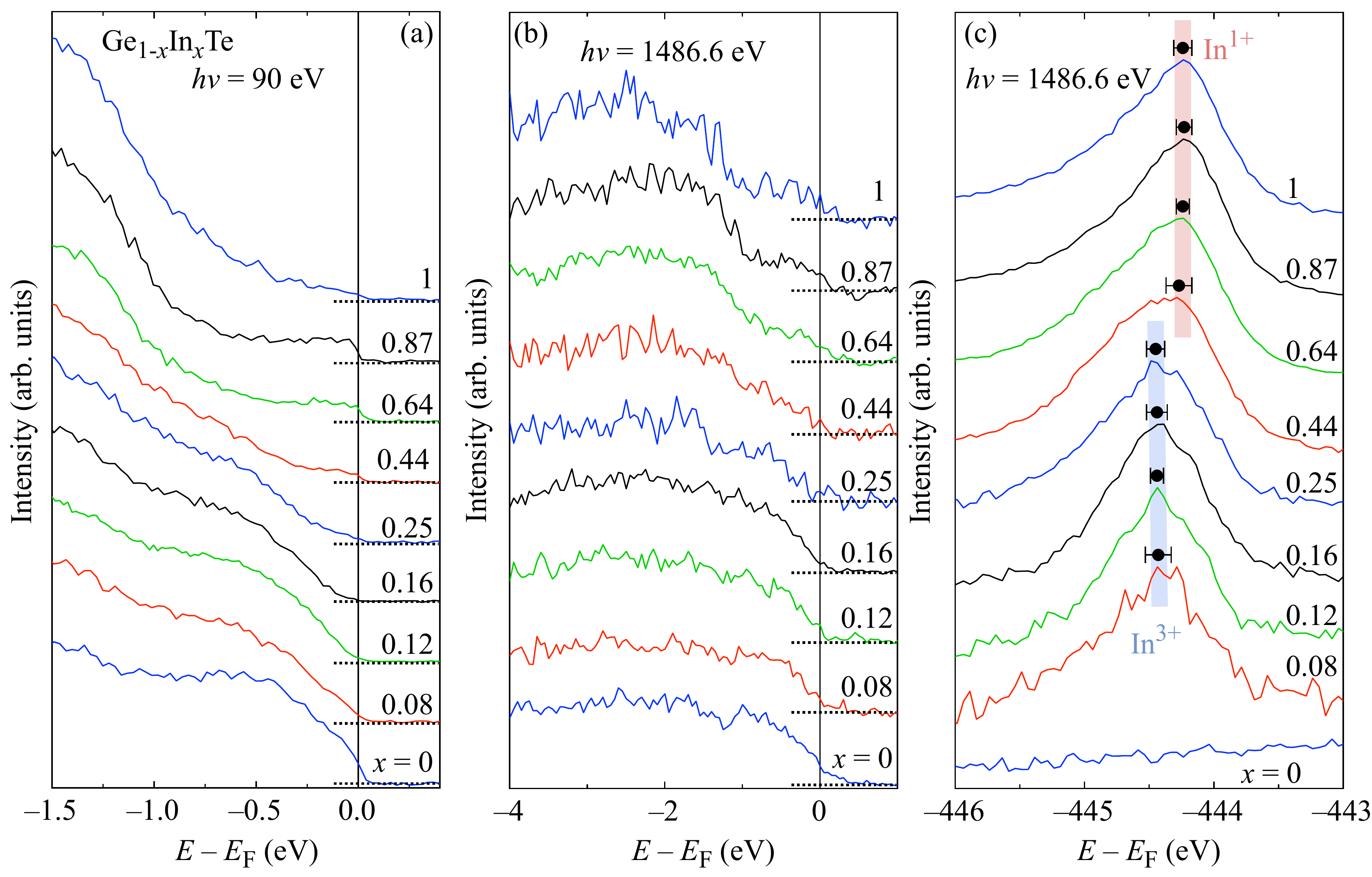}
\caption{Valence-band photoemission spectra of \GIT\ for $0 \leq x \leq 1$ obtained by using (a) a synchrotron-light source ($h\nu = 90$~eV) and (b) an x-ray source ($h\nu =  1486.6$~eV). The dotted lines in both panels indicate the intensity offset. (c) Indium $3d_{5/2}$ core-level photoemission spectra of \GIT\ for $0 \leq x \leq 1$ obtained by using an x-ray source ($h\nu =  1486.6$~eV). The peak energies of the In spectra are indicated by black circles with error bars. The spectra are normalized by the peak heights except for $x = 0$ and shifted with respect to each other for clarity.}
\label{figS6}
\end{figure}

Figures~\ref{figS6}(a) and (b) show the valence-band photoemission spectra of \GIT\ for $0 \leq x \leq 1$ recorded by $h\nu = 90$~eV and $h\nu = 1486.6$~eV, respectively. As already shown in Fig.~4(a,b) of the main text, GeTe exhibits typical $p$-type semiconducting behavior with a degenerate Fermi cutoff at the top of the valence band. Upon increasing $x$ to 0.25, the spectral weight shifts to lower energies and the intensity at the Fermi level gets suppressed, suggesting that electrons are doped, compensating the hole carriers. Above $x = 0.25$, the Fermi cutoff is again clearly observed, indicating metallic behavior. These observations are in good agreement with transport data shown in Fig.~1(d,e) of the main text. We note that the Fermi cutoff of $x = 1$ is smaller than for $x = 0.87$ and 0.64. This may be due to some effect arising from the possible chemical instability of the InTe surface. Indeed, the valence-band spectra recorded by $h\nu = 1486.6$~eV exhibit similar heights of the Fermi cutoff for $x = 0.87$ and 1. The typical probing depth in the $h\nu=90$-eV measurement is $1-5$~\AA, whereas that for the $\sim 1.5$~keV measurement is $4-10$~\AA.\cite{seah79a}

To further clarify the valence state of In, we performed x-ray photoemission spectroscopy. Figure~\ref{figS6}(c) shows the In $3d_{5/2}$ core-level photoemission spectra of \GIT\ for $0 \leq x \leq 1$. For $0.08 \leq x \leq 0.25$, the spectra are nearly $x$-independent, thus indicating that most In ions take the same valence state. At $x = 0.44$, the shape of the spectrum becomes apparently broader with a multi-peak feature. For $0.64 \leq x \leq 1$, the spectra exhibit again a nearly $x$-independent behavior with slightly higher peak energies as compared to $0.08 \leq x \leq 0.25$. The black markers indicate estimated peak positions, where the respective spectral intensities take their maximum values. The peak shifts to a $\sim 200$~meV shallower energy level at around $x = 0.44 - 0.64$, strongly suggesting a change of the valence state. By considering the number of In electrons, we assign the higher (lower) binding energy at $E -\EF$ $\sim -444.4$~eV ($\sim -444.2$~eV) to the In$^{3+}$ (In$^{1+}$) state, respectively, being predominant in the doping range $0.08 \leq x \leq 0.44$ ($0.44 \leq x \leq 1$). We interpret this as a clear signature of a change of the In valence state with $x$. 
%Figure S6

\clearpage

\subsection{Superconductivity in InTe}
\label{SCInTe}
\begin{figure}[h]
\centering
\includegraphics[width=15cm,clip]{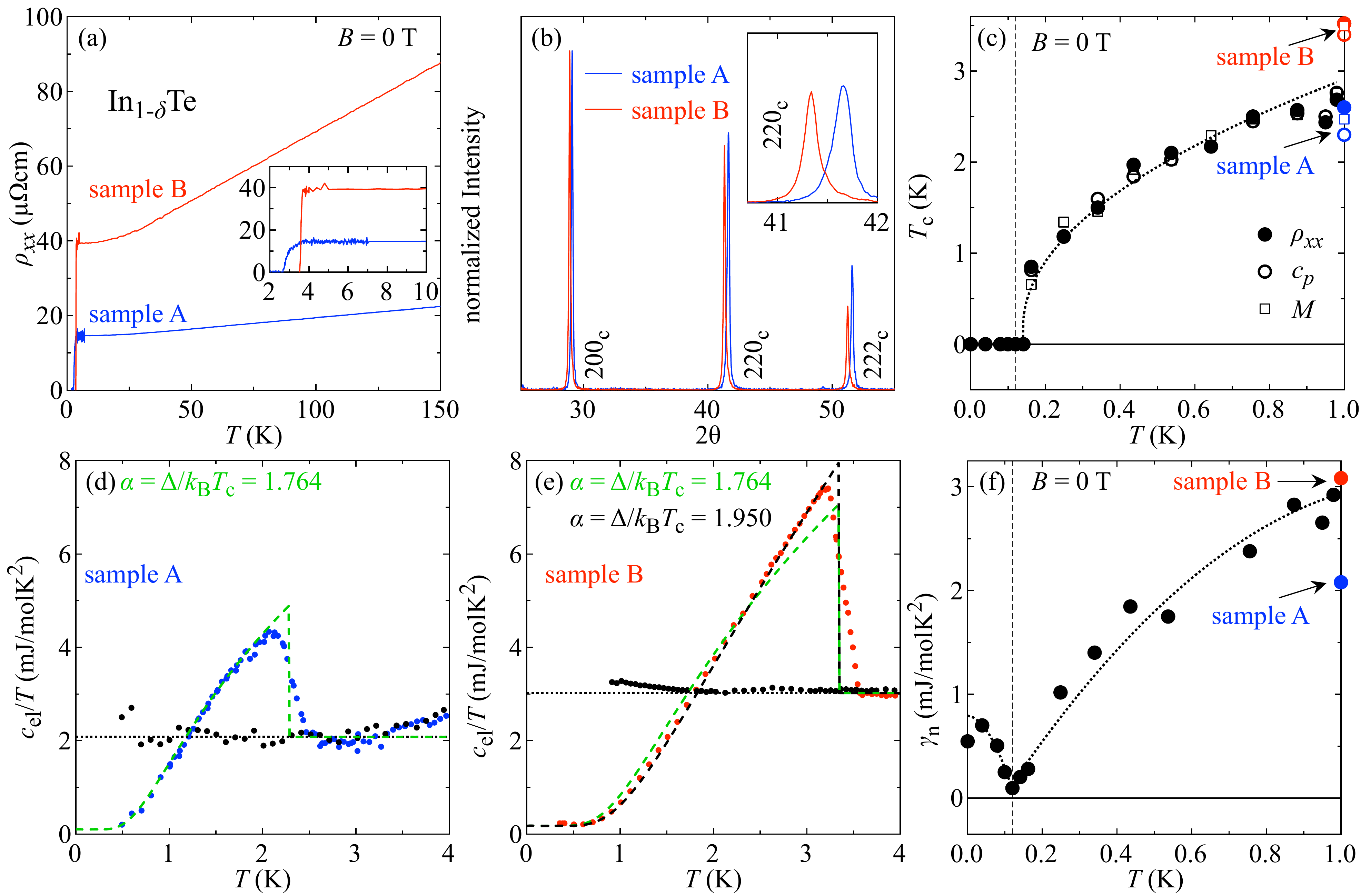}
\caption{Comparison of two samples with nominally $x=1$ and labeled A and B grown according to different high-pressure growth recipes A and B as defined in Section~\ref{grrec}. Data of samples A and B are highlighted in blue and red, respectively. (a) Resistivity \rxx(T). Inset provides an expanded view of the low-temperature region around \Tc. (b) XRD patterns. Inset provides an expanded view around the $220_{\rm c}$ reflection. The difference in the cubic lattice constant is striking. (c) Phase diagram \Tc\ vs $x$. Electronic specific-heat data $\cel/T$ for sample A and B are shown in (d) and (e). The green dashed curves are expected $\cel/T$ in weak-coupling Bardeen-Cooper-Schrieffer (BCS) theory. Apparently this is a very good description in the case of sample A, but not for sample B. A satisfying description of $\cel/T$ of the latter is achieved when increasing the superconducting coupling strength $\Delta/\kB\Tc$ (black dashed curve), see text. In both panels, the electronic specific-heat coefficients \gn\ are indicated by horizontal dotted lines. (f) Phase diagram normal-state specific-heat coefficient \gn\ vs $x$ .}
\label{figS7}
\end{figure}

Figure~\ref{figS7} summarizes data on two InTe samples grown according to the two different growth recipes defined in Section~\ref{grrec} and hence referred to as ``sample A'' (blue data) and ``sample B'' (red data): (a) resistivity, (b) XRD, (c) \Tc\ vs $x$, (d) and (e) temperature-dependent electronic specific-heat $\cel(T)$ data of sample A and B, respectively, and (f) normal-state electronic specific-heat coefficient \gn\ vs $x$. The data shown in Fig.~\ref{figS7}(d) is replotted from an earlier publication of us, see Ref.~\onlinecite{kriener18a}.
Figs.~\ref{figS7}(c) and (f) are also shown in Figs.~3(d) and (e) of the main text. Here data points for sample A are added for comparison. 

\begin{figure}[h]
\centering
\includegraphics[width=8cm,clip]{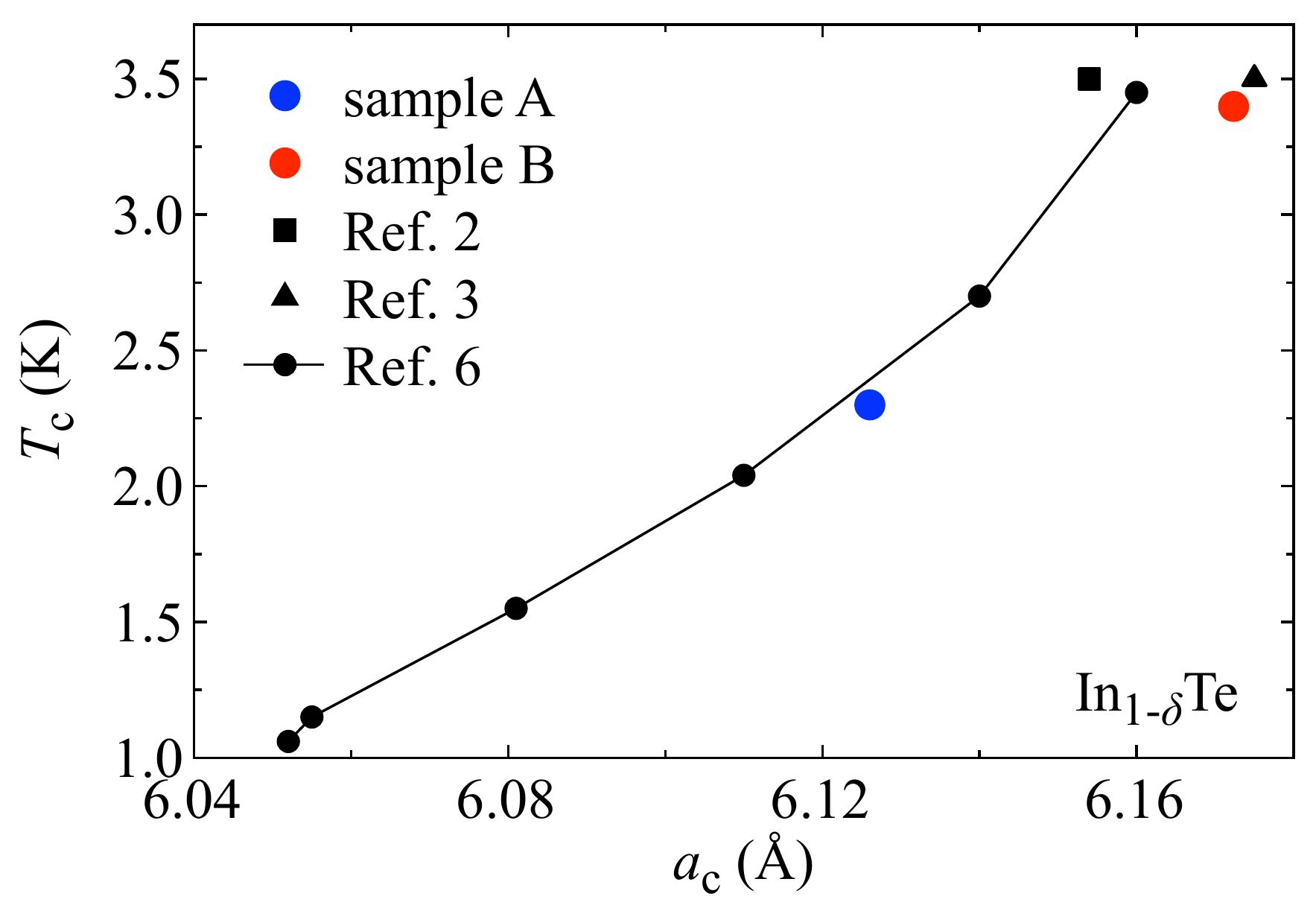}
\caption{Superconducting \Tc\ as function of the cubic lattice constant. The blue and red data point refer to samples A and B, respectively (this work). Black symbols are reproduced from Refs.~\onlinecite{banus63a} (square), \onlinecite{kobayashi18a} (triangle), and \onlinecite{geller64a} (circles).}
\label{figS8}
\end{figure}

Apparently there are several striking differences between these two batches. (i) The \Tc\ of sample B is much higher than that of sample A [by almost 50\%, cf.\ the inset of Fig.~\ref{figS7}(a)]. (ii) As already discussed in Section~\ref{ucv}, the lattice constant of sample B is larger as indicated by the shifted XRD peaks in Fig.~\ref{figS8}(b) and its inset. We note that the peaks of sample B are slightly sharper than those of sample A. The comparison of the electronic specific heats of these two samples shown in Figs.~\ref{figS7}(d) and (e) reveals additional differences between them: (iii) The electronic specific-heat coefficient \gn\ [indicated by the horizontal dotted lines in Figs.~\ref{figS7}(d) and (e)] and hence the density of states is strongly enhanced. Moreover, (iv) while \cel\ can be well described in the standard weak-coupling Bardeen-Cooper-Schrieffer (BCS) scenario, i.e., $\Delta/\kB\Tc=1.764$, this parameter has to be set to 1.950 to reproduce the experimental data of sample B, indicative of more strongly-coupled superconductivity in this sample. By contrast, the Debye temperatures as estimated from specific-heat analyses do not differ much. (v) SEM-EDX analyses yielded a significant difference in the In deficiency in \IT, namely $\delta=0.073$ in sample A and 0.027 in sample B, cf.\ Section~\ref{edx}. Table~\ref{tab1} summarizes these results along with the growth conditions. We note, that there are some counter-intuitive issues. Although sample B has sharper XRD peaks and is much closer to stoichiometry than sample A, the latter exhibits the smaller resistivity. This is also likely the reason why the resistivity of sample B intersects with the resistivity measured for  $x=0.87$ as seen in Fig.~1(e) of the main text. Moreover, the lattice shrinks upon reducing the In content.

According to literature, the \Tc\ of InTe can vary as much as from 1~K to $\sim 3.5$~K.\cite{kobayashi18a,kriener18a,boemmel63a,banus63a,geller64a,geller64b,hulm70a} These changes are attributed to off-stoichiometric In concentrations, i.e., \IT, and hence differences in the lattice constants\cite{geller64a} in agreement with the results of our SEM-EDX analyses as discussed in Section~\ref{edx}. In Fig.~\ref{figS9} we replotted superconducting-transition-temperature data from literature as a function of the cubic lattice constant along with this work's results and find a reasonable agreement. 

As seen in Fig.~\ref{figS3}, the cubic lattice constant is enhanced also in \GIT\ for $x>0.87$, but, even for the sample with nominally $x=0.98$, this does not lead to a similar enhancement of \Tc\ as observed in the case of sample B as can be seen in Fig.~\ref{figS7}(c). However, the electronic specific heat for $x=0.98$ (not shown) is slightly better reproduced when increasing $\Delta/\kB\Tc$ to 1.85. These results may suggest that something unique may be at work in the superconductivity in InTe.
%Figure S7

\begin{table}[b]
\centering
\caption{\textbf{Parameters of two In$_{\bm {1-\delta}}$Te samples grown at different temperatures.}}
\label{tab1}
\begin{tabular}{cccc}
\toprule
Sample                           & A                 & B                 & difference (\%)  \\ \hline \addlinespace[0.75em]
\multirow{3}*{Growth conditions} & 5GPa              & 5~GPa             & -       \\ \addlinespace[0.1em]
                                 & $1200^{\circ}$C   & $600^{\circ}$C    & -       \\ \addlinespace[0.1em]
                                 & 1~h               & 1~h               & -       \\ \addlinespace[0.5em]                             
$\delta$                         & 0.073             & 0.027             & $-37$   \\ \addlinespace[0.5em]
$\Tc(\rho = 0)$                  & 2.60~K            & 3.52~K            & $+35$   \\ \addlinespace[0.5em]
\cp\ midpoint                    & 2.29~K            & 3.35~K            & $+46$   \\ \addlinespace[0.5em]
$a_{\rm c}$                      & 6.126~\AA\        & 6.172~\AA\        & $+0.72$ \\ \addlinespace[0.5em]
\gn\                             & 2.080~mJ/molK$^2$ & 3.020~mJ/molK$^2$ & $+45$   \\ \addlinespace[0.5em]
\gres\                           & 0.100~mJ/molK$^2$ & 0.175~mJ/molK$^2$ & $+75$   \\ \addlinespace[0.5em]
$\Theta_{\rm D}$                 & 168~K             & 162~K             & $-4$    \\ \addlinespace[0.5em]
$\lambda$                        & 0.562             & 0.640             & $+14$   \\ \addlinespace[0.5em]
$\Delta/\kB\Tc$                  & 1.764             & 1.950             & $+11$   \\ \addlinespace[0.5em]

\bottomrule
\end{tabular}
\end{table}

%Figure S8

\clearpage

\subsection{Partial density of states (PDOS) for $\bm {x = 0.12}$}
\begin{figure}[h]
\centering
\includegraphics[width=8cm,clip]{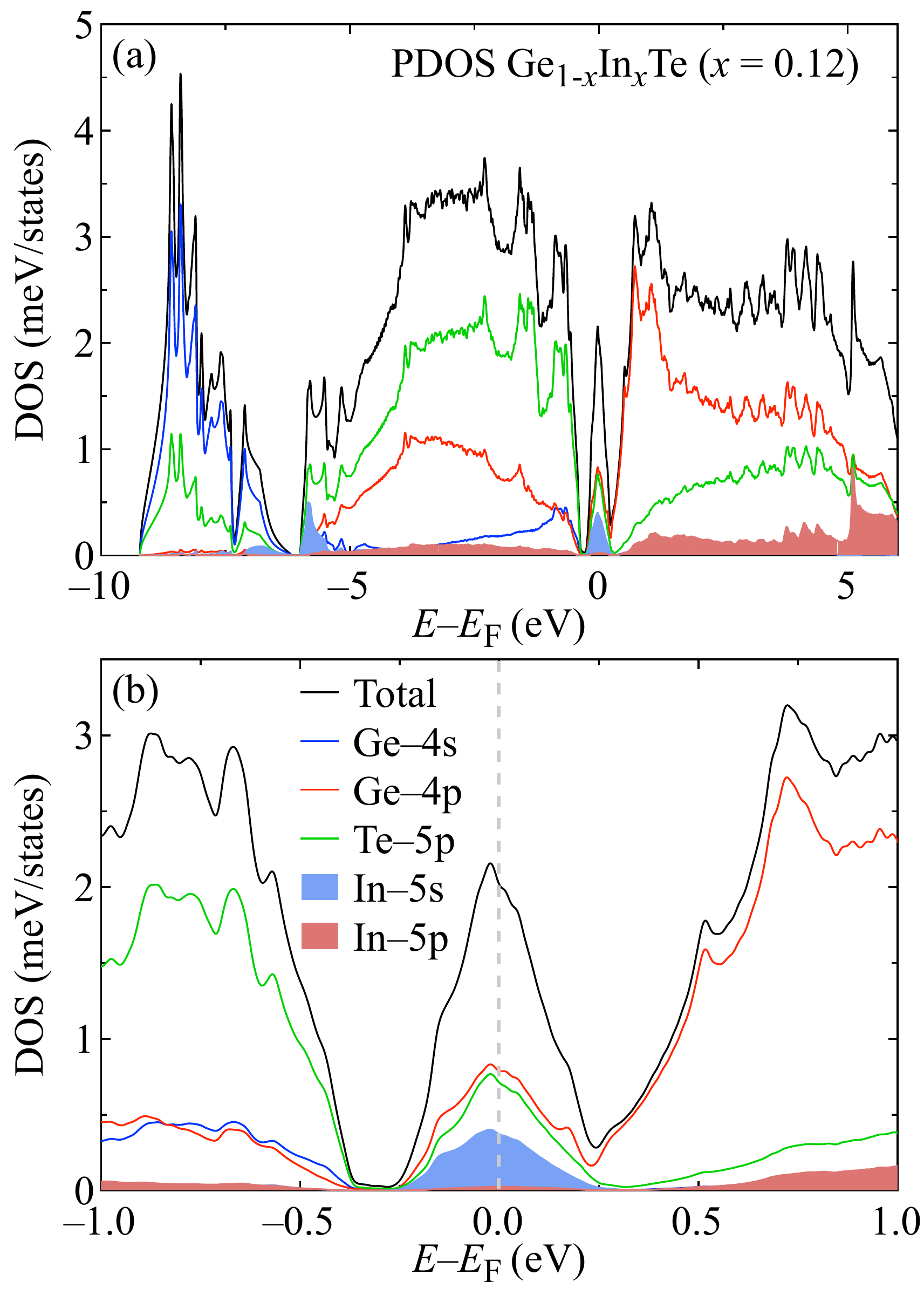}
\caption{(a) Partial density of states for $x = 0.12$; (b) provides an expanded view around the Fermi energy \EF\ indicated by the dashed grey line.}
\label{figS9}
\end{figure}

Figure~\ref{figS9} summarizes calculated partial density of states for $x=0.12$. Panel (b) provides an enlarged view around the Fermi energy \EF. In agreement with the central sketch in Fig.~4(d) of the main text, at this low doping level the In $5s$ and $5p$ bands start to split but are not completely separated yet. The $5s$ state exhibits a larger PDOS at the Fermi level than the $5p$ state which starts to shift mainly upwards above the Fermi energy, but also downwards to hybridize with the Te-$5p$ states. 
%Figure S9

% Bibliography

%\bibliography{/PaperBase/Bibfiles/preload,/PaperBase/Bibfiles/AB-semiconductors,/PaperBase/Bibfiles/Thermoelectrics,/PaperBase/Bibfiles/Superconductivity,/PaperBase/Bibfiles/sonstigePaper,/PaperBase/Bibfiles/Lehrbuch,/PaperBase/Bibfiles/TopolIns,/PaperBase/Bibfiles/SiC-Si-C}